	\documentclass[letter]{emulateapj} 

	\usepackage{natbib}
	\usepackage{graphicx}
	\usepackage{amssymb}
	\usepackage{amsmath}

	\bibpunct{(}{)}{;}{a}{}{,}

	\epsscale{0.5}
	
\begin{document}
	\title{On the polarimetric signature of emerging
	magnetic loops in the quiet-Sun}
	\shorttitle{Stokes~$V$ profiles of emerging magnetic loops}
		
	\author{B.~Viticchi\'{e}\altaffilmark{1}}
	\shortauthors{Viticchi\'{e}}
	
	\altaffiltext{1}{ESA/ESTEC RSSD, Keplerlaan 1, 2200 AG Noordwijk, Netherlands}
	\email{Bartolomeo.Viticchie@esa.int}

\begin{abstract}
The abundance of Stokes~$V$ profiles dominated
by one lobe at the locations of emergence of $\Omega$-shaped
magnetic loops is evaluated. The emergence events were found in \textit{Hinode} SOT/SP
time-sequences of quiet-Sun regions. Such a study has the aim of confirming
a prediction based on the basic geometrical and physical
properties of emerging magnetic loops: Stokes~$V$ profiles dominated
by one lobe are possibly the main polarimetric signature of these structures.
In agreement with this prediction, $47$\% of the Stokes~$V$ profiles analyzed
has an amplitude asymmetry $\vert \delta a \vert>0.3$, while in the quiet-Sun the abundance is
of about $30$\%. This excess with respect to the quiet-Sun is found consistently for
whatever value of the threshold on the amplitude asymmetry.
Such a result proves the goodness of the physical scenarios
so far proposed for the interpretation of loop emergence events and
may prompt the use of Stokes~$V$ profiles dominated by one lobe as
a new proxy for their identification
in observations with a good spectral sampling.
\end{abstract}

	\maketitle

\section{Introduction}
\label{Intro}
	The spectro-polarimeter SOT/SP \citep[][]{Lit01} aboard the JAXA
	mission Solar-B \citep[\textit{Hinode},][]{Kos07} allows one to perform
	high spatial resolution ($0.3$~arcsec), and high polarimetric sensitivity
	($10^{-3}$ the quiet-Sun continuum intensity) spectro-polarimetric
	observations of the solar photosphere in the \ion{Fe}{1}~$630$~nm lines.
	Since its launch in $2006$, SOT/SP provides to the
	solar community the conditions to achieve a breakthrough in
	the investigation of the quiet-Sun magnetism.
	
	Recently, great attention has been dedicated
	by the solar community to the study of
	magnetic field emergence events in SOT/SP data. These
	were firstly pointed out either as small scale
	magnetic loops \citep[loop cases;][]{Cen07}, or magnetized
	emerging granules with a single magnetic polarity \citep[unipolar cases;][]{OroS08}.
	Latterly, in \citet[][]{MarG09}, an extensive analysis of approximately $50$
	emerging loops anchored to the solar photosphere ($\Omega$ loops) was presented.
	In \citet[][]{MarG10} both the topology and the dynamics
	of an emerging $\Omega$ loop were derived. A similar study
	was presented in \citet[][]{Ish10}.
	More studies of magnetic field emergence events
	can be found in the literature: the ones of \citet[][]{MarG07}, and \citet[][]{Gom10}
	performed on data from TIP \citep[][]{Col99}, and the ones of \citet[][]{Gug11}, and
	\citet[][]{Pal11} performed on data from IMaX \citep[][]{MarP11}.
	A complete understanding of the emergence of magnetic fields
	in the quiet-Sun could considerably improve our knowledge of the solar
	photosphere \citep[see the references in the introduction of][]{MarG09}.
	
	In spite of the rich literature on the emergence of magnetic fields in the
	solar photosphere, there is still a lack of knowledge about
	the typical polarimetric signatures associated to these events. In this work, the first analysis of the
	shapes of Stokes~$V$ profiles measured at the locations of
	$\Omega$ loop emergence events in SOT/SP data is presented. The main goal of this
	study is to confirm the following prediction that can be
	outlined by considering the basic geometrical and physical properties of emerging $\Omega$ loops:
	Stokes~$V$ profiles dominated by one lobe are possibly the main polarimetric signature
	associated to such events in the quiet-Sun.
	
	The paper is organized as
	follows: the polarimetric signatures which are expected to be observed
	at the locations of an emerging $\Omega$ loop are described in Sect.~\ref{Pred};
	the dataset and the analysis method adopted	are presented in Sect.~\ref{Data&Meth};
	the results of the analysis confirming the prediction made
	in Sect.~\ref{Pred} are presented and discussed in Sect.~\ref{ResDisc};
	the conclusions are outlined in Sect.~\ref{Conc}.
	
\section{Polarimetric signatures of an emerging $\Omega$ loop}
\label{Pred}
\begin{figure}[!t]
		\centering
		\includegraphics[width=6cm]{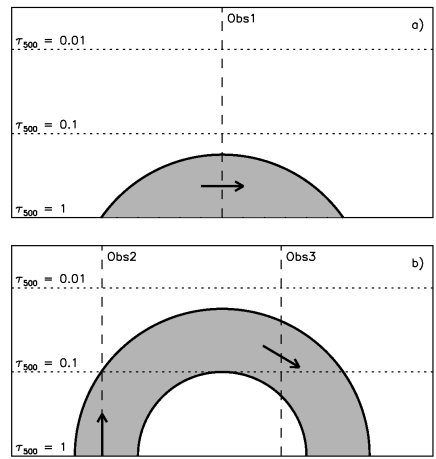}
		\caption{Cartoon drawing of a magnetic $\Omega$ loop
		(shaded area) emerging above the solar photosphere.
		Upper panel: initial phase
		producing the linear polarization signature.
		Lower panel: the following phase, with the
		loop standing above the photosphere, producing
		the two circular polarization signatures.
		The location of the photospheric layer is marked
		by the average $\tau_{500}=1$ level.
		The upper photosphere layers are marked by the
		average $\tau_{500}=0.1,~\mathrm{and}~0.01$ levels (horizontal
		dotted lines). The dashed vertical lines show
		three LOS examples crossing the loop structure
		at different instants and locations.
		The arrows crossing the LOS represent
		the direction of the magnetic field vector.\label{fig1}}
	\end{figure}

	Figure~\ref{fig1} shows a cartoon drawing of a $\Omega$ loop
	which is emerging in a field-free environment at the disk
	center. More precisely, Fig.~\ref{fig1} represents the two phases
	which allow one to identify a loop emergence event in
	time-sequences of polarimetric data. In this
	figure, the different layers of the solar photosphere are
	marked with average values of the optical depth at $500$~nm ($\tau_{500}$).
	In the initial phase (\texttt{a} panel in Fig.~\ref{fig1}) the top of the
		loop, dominated by transversal magnetic fields with
		respect to the line-of-sight (LOS) of an observer (e.g. \texttt{Obs1}),
		enters the photosphere producing a linear polarization signature
		\citep[usually on top of granules, e.g.][]{Ish10}.
	In the following phase (\texttt{b} panel in Fig.~\ref{fig1}), the loop emerges
		above the photosphere and two circular
		polarization signatures with opposite polarities,
		produced by the longitudinal fields with respect to the LOS,
		are observed (e.g. by the observer \texttt{Obs2}).
	The polarization signatures associated to these two phases of
	a magnetic loop emergence event are commonly used as proxies to
	identify emerging loops in polarimetric observations
	of the quiet-Sun (see the references in Sect.~\ref{Intro}).

	However, an in-depth understanding of $\Omega$ loop emergence events
	requires dealing with the details of polarization measurements.
	In \citet[][Fig.~8]{Ish10}
	the authors derived the physical properties of an emerging $\Omega$ loop
	by collecting one-dimensional (1D)
	slices retrieved from inversions of polarization
	measurements. It is worth pointing out that
	the authors had to deal with Stokes~$V$
	profiles with strong asymmetries: two Stokes~$V$ profiles
	dominated by one lobe were observed in two of the pixels of the
	loop \citep[see Fig.~3 of][]{Ish10}.
	In the following it is shown that it is possible to predict that this
	kind of profiles are possibly the main polarimetric
	signature associated to emerging $\Omega$ loops.
	
	The LOS \texttt{Obs2}, crossing one of the
	loop's foot-points in Fig.~\ref{fig1}, passes
	first through a field-free layer which fills the region
	$\tau_{500}=0.01-0.1$, and then through a
	magnetized layer which fills the region $\tau_{500}=0.1-1$.
	If the location of the foot-point crossed by \texttt{Obs2}
	is an up-flow (as usual in emergence events),
	the plasma velocity along the LOS  in the region $\tau_{500}=0.1-1$ is
	negative and of the order of several kilometers per second,
	while the same quantity in the region $\tau_{500}=0.01-0.1$ is
	expected to be considerably smaller in absolute value \citep[see e.g.][]{VitNVit11}.
	From the extensive literature on the formation of strongly asymmetric
	Stokes~$V$ profiles in Visible spectral lines from 1D
	stratifications one can conclude that an atmospheric
	configuration like the one described above can give rise to
	a Stokes~$V$ profile dominated by the blue lobe under very different conditions:
	for either sharp or smooth transitions between the magnetized region
	and the field-free one, for either strong or weak magnetic field regimes,
	for either vertical or inclined magnetic fields with respect to the LOS.
	Conversely, if the foot-point is located in a down-flow, a profile dominated
	by the red lobe is very likely to be formed \citep[][]{VitNVit11,SaiD11}.
	Similar arguments can be used to predict that the
	atmospheric properties along the LOS \texttt{Obs3} can
	give rise to Stokes~$V$ profiles dominated by one lobe as well \citep[][]{GroD00,VitNVit11,SaiD11}.
	Figure~3 of \citet{Ish10} offers one a fast verification
	of the prediction outlined above. In Sect.~\ref{ResDisc}, much stronger arguments in favour of
	the connection between $\Omega$ loop emergence events and the observation
	of Stokes~$V$ profiles dominated by one lobe are presented.
	\begin{figure*}[!t]
		\centering
		\includegraphics[width=10cm]{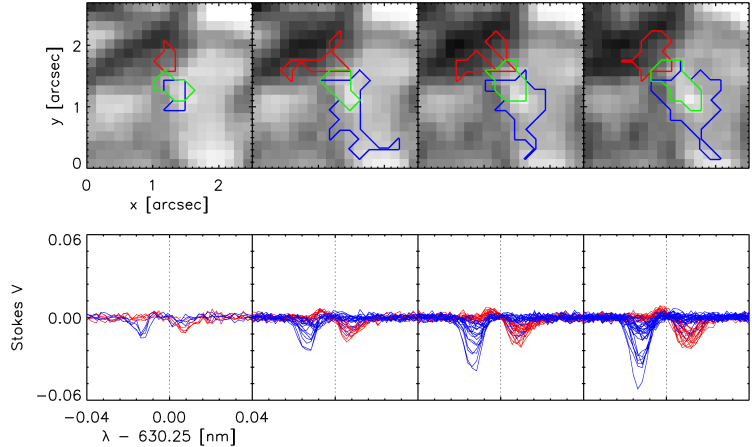}
		\caption{An example of a $\Omega$ loop emergence event extracted
		from the analyzed time-sequences. Upper row: sub-sequence of Stokes~$I$
		continuum sub-fields centered around the location of the emergence event.
		The contours mark the positions of the positive circular
		polarization signature (red contours), the negative circular polarization
		signature (blue contours), and the linear polarization
		signature (green contours). These allow one to point out the emergence
		event according to the criteria used in \citet{MarG09}.
		The contour regions have areas larger than four pixels of \textit{Hinode}~SOT/SP and
		max$(\vert Q \vert, \vert U \vert)>8\times10^{-3}$ and
		max$(\vert V \vert)>6\times10^{-3}$ (see Sect.~\ref{Data&Meth}).
		Lower row: Stokes~$V$ profiles with $\vert \delta a \vert>0.3$ from the pixels in the
		locations of the two circular polarization signatures marked by the blue
		and red contours in the upper row. The colors encode the polarimetric signature from which the profiles
		were picked.\label{fig2}}
	\end{figure*}
	\begin{figure}[!t]
		\centering
		\includegraphics[width=6cm]{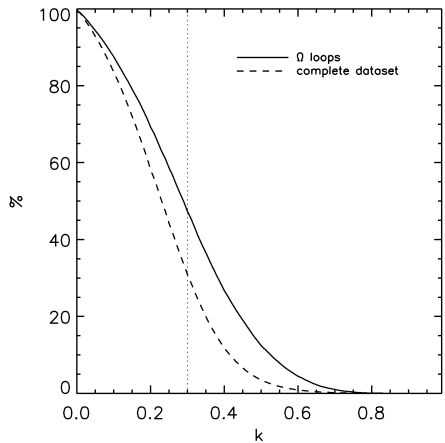}
		\caption{Abundances of Stokes~$V$ profiles with $\vert \delta a \vert>k$
		as functions of $k$. The solid line shows the abundances derived
		at the locations of emergence of magnetic loops; the dashed line shows the ones
		derived from the complete quiet-Sun dataset of all the time-sequences when considering
		the profiles with $\mathrm{max}(\vert V \vert)>0.006$ (consistent with Sect.~\ref{Data&Meth}).
		The dotted vertical line marks the threshold value here used to evaluate the
		abundances, i.e. $k=0.3$.\label{fig3}}
	\end{figure}
	
	The polarimetric signatures produced by the emerging $\Omega$ loop in Fig.~\ref{fig1}
	are assumed to have dimensions
	larger than the resolution element of modern
	spectro-polarimetric observations, i.e. the loop
	structure is resolved\footnote{For a discussion on the polarimetric
	signature produced by unresolved magnetic loops
	refer to \citet{Ste00}.}. This assumption does
	not imply that the magnetic fields of the loop are
	completely resolved, it rather implies that a
	dominant $\Omega$-shaped magnetic structure can be pointed out
	from the data.
	This structure is the one which is
	considered to be dominant in each pixel of the
	observed loops.
	For this reason one can use the properties
	of 1D stratifications to predict the asymmetries of Stokes~$V$
	profiles at the locations of emerging magnetic loops.
\section{Dataset and analysis method}
\label{Data&Meth}
	In order to have a large statistics of Stokes~$V$
	profiles from the locations of emerging magnetic loops
	one can use the time-sequences analyzed in \citet{MarG09}.
	This is so far the
	study with the richest record of emergence events of
	$\Omega$ loops in the quiet-Sun.
	In detail, all the time-sequences reported
	in Table~1 of \cite{MarG09}, with exception made for
	the sequence of September $26$th (due to
	problems in downloading the dataset from the \textit{Hinode}
	archive), were analyzed. One can refer to both Table~1 and the second paragraph of Sect.~2
	of \cite{MarG09} for the description of the datasets.
	
	As reported in Sect.~\ref{Pred}, a magnetic $\Omega$ loop
	can be identified in polarimetric data as ``a linear polarization signature
	flanked by circular polarization signatures with
	opposite polarities'' \citep[e.g.][]{MarG09}.
	In the analysis here presented, those events in which
	the linear and circular polarization signatures had,
	at least for one wavelength, max$(\vert Q \vert, \vert U \vert)>8\times10^{-3}$ and
	max$(\vert V \vert)>6\times10^{-3}$ in each pixel, respectively, were considered.
	Besides this, a minimum area of four pixels for each polarization
	signature was required\footnote{Strong
	and well-resolved polarization signatures with a maximum Stokes~$V$
	amplitude above three times the
	polarimetric sensitivity of the observations, i.e. $1.7\times10^{-3}$ in units
	of the average continuum intensity
	\citep[see][]{MarG09}, were considered. The higher threshold for the individuation of the linear
	polarization signature allows one to point out sound
	cases.}.
	Each emergence event was taken out of the
	time-sequences by picking out a sub-sequence
	of sub-fields (of $\simeq2.7\times2.7$~arcsec$^2$)
	around the location where the emergence
	was found to take place. Spurious polarimetric signals due to magnetic
	fields in the selected sub-fields were then carefully removed
	along the whole sub-sequence by visual inspection.
	
	An example of emergence event extracted from the analyzed sequences
	is shown in the upper row of Fig.~\ref{fig2} in which both the linear and
	the circular polarization signatures that allow one to point out the $\Omega$
	loop were marked (contours). A total of $40$ events were found in the analyzed sequences;
	these presented a large variety of properties which were investigated in detail in
	\citet{MarG09}. Here the abundance of Stokes~$V$ profiles dominated by one lobe
	at the locations of the circular polarization signatures of the loops was studied.
	
	The lower row of Fig.~\ref{fig2} shows a fraction of
	the Stokes~$V$ profiles at the locations of the circular polarization
	signatures of the emergence event in the upper row. In detail, among the
	Stokes~$V$ profiles observed at the locations of the
	circular polarization signatures of the $\Omega$ loop, the ones with
	$\vert \delta a \vert=\vert \frac{\vert a_b \vert-\vert a_r \vert}{\vert a_b \vert+\vert a_r \vert}\vert >0.3$
	were represented\footnote{In the formula, $a_b$ ($a_r$) is the amplitude of the blue (red) lobe.}.
	$\delta a$ is the amplitude asymmetry of a Stokes~$V$: the higher
	the $\vert \delta a \vert$ the more the Stokes~$V$ is dominated by one lobe.
	
	One can understand the formation the Stokes~$V$ profiles in Fig.~\ref{fig2} by
	referring to Sect.~\ref{Pred} and Fig.~\ref{fig2} (\texttt{Obs2}).
	The signature with the positive polarity, mostly located in a down-flow, is
	characterized by Stokes~$V$ profiles dominated by the red lobe. The negative one,
	located on a bright granule, is characterized by profiles dominated by the
	blue lobe. The observations of these polarimetric signatures were predicted
	in Sect.~\ref{Pred}.
	
	By exploiting all the $40$ $\Omega$ loop emergence events
	individuated in the analyzed time-sequences one can analyze
	more profiles and, eventually, evaluate the abundance of Stokes~$V$
	profiles dominated by one lobe at
	the locations of such events. To do this, all the
	Stokes~$V$ profiles from the circular polarization signatures
	which allow one to identify the emergence events were collected;
	these were $12256$ profiles. Those profiles with
	$\vert \delta a\vert>0.3$ were the ones considered
	to be dominated by one lobe (see e.g. Fig.~\ref{fig2}).
\section{Results and Discussion}
\label{ResDisc}
	The Stokes~$V$ profiles with $\vert \delta a \vert>0.3$
	make up $5799$ of the $12256$ of the whole archive
	(i.e. $47$\%). This means that nearly one in every two Stokes~$V$ profiles
	observed at the locations of emerging magnetic loops is dominated either by
	the blue lobe or by the red lobe. Such a result
	confirms the prediction made in Sect.~\ref{Pred}, and allows one
	to put forward Stokes~$V$ profiles dominated by one lobe as a new proxy for emerging
	$\Omega$ loops in polarimetric observations with good spectral sampling.
	Even though the abundance here derived stemmed from
	an arbitrary threshold on the amplitude asymmetry of the analyzed circular
	polarization signals, the plenitude of Stokes~$V$ profiles dominated by one lobe
	at the locations of emerging magnetic loops can be proved in
	an alternative way.	Figure~\ref{fig3} shows the abundances
	for different values of the threshold ($k$ in Fig.~\ref{fig3}) on the amplitude
	asymmetry in both the sample described at the end of Sect.~\ref{Data&Meth} 
	and the complete quiet-Sun dataset of all the time-sequences. On the one hand, the
	result reveals that the abundance strongly depends on the threshold value.
	On the other hand, it shows that there is a considerable excess of profiles
	with large $\vert \delta a \vert$ at the locations of loop emergence events
	with respect to the quiet-Sun. This result does not depend on the value
	of the threshold, and further supports the prediction made in Sect.\ref{Pred}.
	
	Polarimetric observations of loop emergence events performed with other
	instruments do not present the same abundance of extremely asymmetric profiles.
	\citet{Gom10} presented a study of the shapes of Stokes~$V$
	profiles observed in the two infra-red (IR) \ion{Fe}{1} lines at $1565$~nm with
	TIP \citep[][]{Col99} at $1$~arcsec spatial resolution. The profiles
	they showed in Fig.~2 are anti-symmetric profiles. Such a lack of
	asymmetry can be understood by referring to
	\citet{GroD89} in which the authors explained the different conditions
	which produce strong asymmetries in Stokes~$V$ profiles in the Visible and
	in the IR.
	In \citet{Gug11} the authors pointed out the observation of asymmetric profiles
	in one emergence event observed with IMaX
	\citep[][]{MarP11} in the \ion{Fe}{1}~$520$~nm line.
	A detailed analysis of the profiles observed by IMaX at the locations
	of emergence events would be important to complement the study
	presented here.
		
	Finally, the results shown here allow one to complement the knowledge about
	the origin of Stokes~$V$ profiles dominated by one lobe observed by SOT/SP in the quiet-Sun.
	In detail, in \citet{VitNVit11} the authors reported on the observation of
	such profiles at the borders of network patches in the dataset of \citet{Lit08},
	while in \citet{OroS08} these were found at the locations of unipolar emergence events.
	
\section{Conclusions}
\label{Conc}
	The first study of the shapes
	of Stokes~$V$ profiles observed at the locations of
	emerging magnetic $\Omega$ loops found in SOT/SP time-sequences
	of the quiet-Sun is presented. It focuses on those loops which can
	be pointed out adopting the standard method based on the
	detection of both linear and circular polarization
	signatures in \textit{Hinode} time-sequences of the quiet-Sun \citep[see e.g.][]{MarG09}.
	$47$\% of the profiles observed at the locations
	of such events have an amplitude asymmetry 
	$\vert \delta a \vert>0.3$, i.e. are dominated by either
	the blue lobe or the red lobe. Such a value
	quantifies the excess of the profiles dominated by one lobe
	at the locations of magnetic loop emergence events with respect to
	their abundance in the complete quiet-Sun dataset of all
	the time-sequences.
	
	This result confirms the prediction of
	the Stokes~$V$ profiles that are expected to be
	observed at the locations of emerging loops outlined in Sect.~\ref{Pred}.
	The different LOS crossing the structure of a loop
	are expected to define 1D
	stratifications which are known to produce
	Stokes~$V$ profiles dominated by one lobe
	under very different conditions in Visible spectral lines
	\cite[][]{GroD89,Ste00,GroD00,VitNVit11,SaiD11}.
	The large abundance of these profiles at the locations of
	loop emergence events confirms the goodness
	of the physical scenarios so far proposed to interpret
	such events \citep[e.g.][]{MarG09} and may prompt the use
	of Stokes~$V$ profiles dominated by one lobe as proxies to individuate
	emerging magnetic loops in spectropolarimetric observations
	with good spectral sampling.
	
	The work presented here must be considered
	as a starting point for more refined studies of emergence events
	which, for example, might be aimed to: point out various emergence
	configurations, understand the temporal evolution of the Stokes~$V$
	profiles as a loop emerges/evolves, measure the siphon-flows into the loops.
	
\acknowledgements
	The author is grateful to the anonymous referee for the
	comments which allowed to considerably improve the whole
	manuscript. The author thanks C.~Fisher, M.~Mart\'inez Gonzalez,
	J.~S\'anchez Almeida, and L.~Bellot Rubio, for the useful
	discussions on the topic and comments about the paper.
	The author dedicates this work to his beloved Sara. The data used here
	were acquired during the \textit{Hinode} Operation
	Plan 14, entitled ``\textit{Hinode}-Canary Island campaign''.
	\textit{Hinode} is a Japanese mission developed and launched by ISAS/JAXA,
	collaborating with NAOJ as a domestic partner, NASA and STFC (UK) as
	international partners. Scientific operation of the \textit{Hinode} mission is
	conducted by the \textit{Hinode} science team organized at ISAS/JAXA. This team
	mainly consists of scientists from institutes in the partner countries.
	Support for the post-launch operation is provided by JAXA and NAOJ (Japan),
	STFC (U.K.), NASA, ESA, and NSC (Norway).

\end{document}